# Emerging Ideas in Nanocantilever based Biological Sensors

Ankit Jain[*] and Muhammad Ashraful Alam[#]

School of ECE, Purdue University, West Lafayette, IN, USA, 47907

e-mail: [*]jankiit@gmail.com, [#]alam@purdue.edu

*Abstract*

In this review article, we focus on emerging nanocantilever based biological sensors and discuss the response of nanocantilevers towards bio-molecules capture. The article guides the reader through various modes of operation (e.g., static or dynamic) to detect the change in cantilever's characteristics (e.g., mass, stiffness, and/or surface stress) due to adsorption of bio-molecules on cantilever surface. First, we explain the classical linear resonant mode mass sensors and static stress based sensors. The effect of operating the cantilever in nonlinear regime is then illustrated through examples of bifurcation based mass sensors and electromechanical coupling based Flexure-FET biosensors. We believe that a new class of nonlinear sensors, with their extraordinary sensitivity towards bio-molecules capture, could be the potential candidate for low cost point-of-care applications.

## 1. INTRODUCTION

Detection of biological molecules e.g., viruses, proteins, DNA, etc., is essential for food safety, early warning of biological attack, early stage diagnosis of cancer, and genome sequencing. Nanoscale devices are widely regarded as a potential candidate for ultra-sensitive, low-cost, label-free detection of bio-molecules and are considered as a technology alternative to the existing chemical or optical detection schemes. Label-free schemes detect bio-molecules using their intrinsic properties, e.g., size, mass or charge of a molecule, instead of using extrinsic optical or magnetic labels attached to the target molecule. Among the various label-free technologies, significant research has focused on developing ultra-sensitive biological sensors based on nanocantilevers [1], [2].

The use of a cantilever as a sensor dates back to 1943 when Norton proposed a hydrogen gas sensor based on a cantilever [3]. The opportunity to develop the cantilever as a highly sensitive biosensor, however, had to wait the invention and wide-spread adoption of atomic force microscope (AFM) [4]. An AFM measures the forces between the tip of a cantilever and the sample surface using the tip deflection (contact mode AFM) or changes in the resonance frequency of a vibrating cantilever (dynamic mode AFM). As we will see in Sec. 2, nanocantilever based biosensors operate in a closely related principle, where interaction with biological molecules changes the bending (static mode) or resonance frequency (resonant mode) of the cantilever [5]. Note that these mechanical sensors offer an advantage of detecting both charged as well as neutral bio-molecules; in contrast, their electronic counterpart, e.g., ISFETs (Ion Sensitive Field Effect Transistors [6]) can only detect charged molecules. Today, cantilever-based devices find applications in broad range of fields such as communication, computation, optics, scanning probe microscopy, and sensing. Specific examples include RF-MEMS capacitive/ohmic switches [7], varactors, tunable oscillators [8], NEMS relays [9], NEMFET [10], deformable mirrors [11], displays [12], accelerometers [13], and chemical/gas sensors. In this article, however, we only focus on the use of cantilever as a biological sensor, and discuss both classical and emerging modes of biosensor operation.

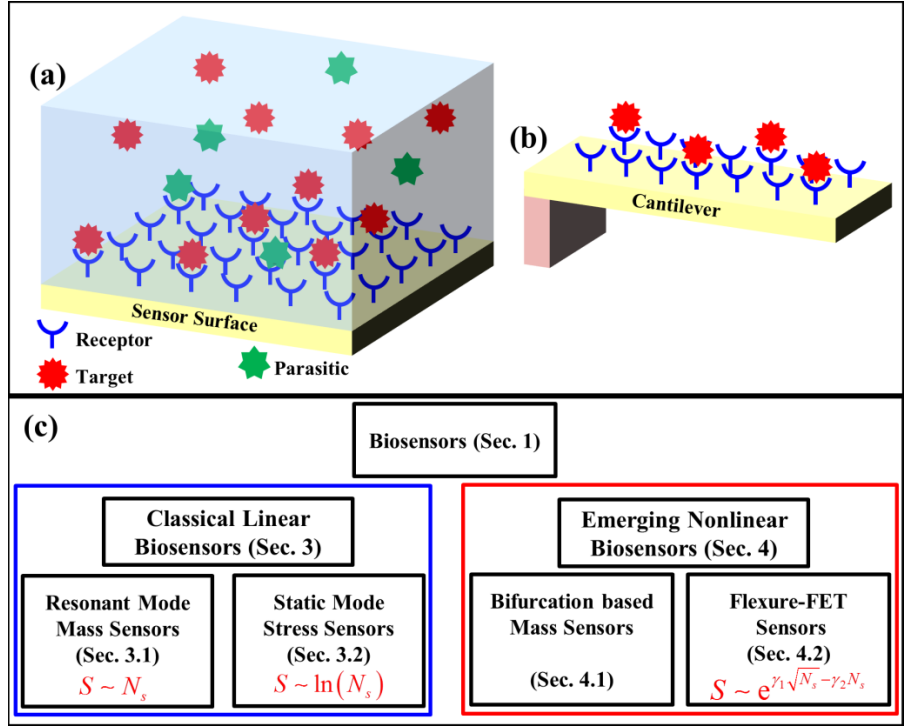

Fig.1: **(a)** Schematic of a generic biological sensor operating in fluidic environment. The receptor, target and parasitic molecules are also shown. **(b)** Drawing of a nanocantilever in which capture of target molecules changes the cantilever's static or dynamic response through change in the mass, stiffness and/or surface stress. **(c)** A chart showing various linear and nonlinear cantilever biosensors to be discussed in this article. $S$ is the sensitivity of respective biosensors and $N_s$ is the areal density of captured bio-molecules on the cantilever surface.

Figures 1(a)-(b) show the schematic of a sensor surface operating in a fluidic environment. The sensor surface is functionalized with receptor molecules so that it can subsequently conjugate to the target molecules (without any optical or magnetic labels) contained in the fluid. For example, if the target is an antibody, then receptor is corresponding antigen, or if the target is a DNA base (e.g., A), then receptor is its conjugate base (e.g., C). The specific binding of the target and receptor molecules changes the sensor characteristics and the change is measured as a signature of detection. For example, the binding may induce an additional surface stress or may change the mass or stiffness of the cantilever, resulting in bending or change in the resonance frequency [5].

There are three key metrics of any sensing scheme (electronic or mechanical), namely, response time [14], selectivity [15], and sensitivity [16]. *Response time* is the time required to capture a certain number of target molecules to get a detectable output signal. Interestingly, response time depends on the geometry of the sensor surface and cylindrical geometry allows the smallest response time [14]. On the other hand, *selectivity* is associated with the problem of non-specific binding, i.e., binding of parasitic molecules with receptor molecules, producing a "false-positive" signal. In a highly selective sensor, receptor molecules should only bind to the target molecules and not to any other molecules in the solution. Finally, *sensitivity* can be defined in number of ways; in general, it is measured as the change in the sensor characteristics (e.g., resonance frequency of a cantilever [5] or drain current of ISFET [17]) in response to the capture of a given number of target molecules. Note that, response time and selectivity of a sensor do not depend on the sensing scheme, whereas sensitivity depends on the sensing scheme. Therefore, in this article, we only discuss sensitivity related issues of nanocantilever based biological sensors. Note that, sensitivity of the sensor depends whether it is operated in linear or nonlinear regime. Figure 1(c) summarizes all linear and nonlinear cantilever biosensors to be discussed in the following sections.

The rest of the article is organized as follows. In section 2, we discuss the spring-mass model of cantilever and illustrate classical mass and stress based sensors in section 3. We then present the emerging nonlinear biosensors like bifurcation based mass sensors and Flexure-FET in section 4. We finally conclude in section 5.

## 2. CANTILEVER BASED SENSORS AS SPRING MASS SYSTEM

The static as well as dynamic response of cantilever based sensors is best described by Euler-Bernoulli beam equation [18]. In this article, we however use a lumped parameter, spring-mass system (Fig. 2(b)) of a cantilever to illustrate its key features. The equation of motion of lumped parameter spring-mass system is given by-

$$m\frac{d^2y}{dt^2} + \frac{m\omega_0}{Q}\frac{dy}{dt} - k(y_0 - y) - k^{'}(y_0 - y)^3 = F_{ext}, \qquad (1)$$

where $m$ is the effective mass of the cantilever, $y$ is the position of vibrating cantilever, $t$ is time, $\omega_0$ is the natural frequency of cantilever, $Q$ is the quality factor, $k$ is the effective spring constant of the cantilever such that $\omega_0 = \sqrt{k/m}$, $y_0$ is the position of the cantilever in its rest position, and $k'$ is the constant associated with cubic nonlinearity of spring. Note that, $k = \frac{\alpha E W H^3}{(1-\nu)L^3}$ is the spring constant of the cantilever where $\alpha$ is a geometrical factor, $E$ is the Young's modulus, $\nu$ is the Poisson's ratio, $W$ is the width, $H$ is the thickness, and $L$ is the length. $F_{ext}$ is the external force acting on the cantilever, e.g., surface forces, electrostatic forces, etc. Historically, the cantilever based sensors have been operated without applying any external force (i.e., $F_{ext} = 0$) and in linear response regime i.e., $k' \approx 0$ (section 3). We will explore the nonlinear $F_{ext} \neq 0$ & $k' \neq 0$ operation in section 4.

## 3. CLASSICAL LINEAR BIOSENSORS

### 3.1 *Resonant mode mass sensors*

In resonant mode sensing, vibrating nanocantilever can be used as a microbalance and bio-molecules can be detected by observing the change in dynamic response of the cantilever [1], [5]. Fundamentally, dynamic response of a cantilever is governed by its resonance frequency $f_0$ that is given by (using Eq. 1 with $k' = 0$ and $F_{ext} = 0$)-

$$f_0 = \frac{\omega_0}{2\pi} = \frac{1}{2\pi}\sqrt{\frac{k}{m}}. \qquad (2)$$

Once the target molecules are captured, change in $m$ (and/or $k$) shifts the resonance frequency ($\Delta f$) to indicate the capture of bio-molecules. Experimentally, resonance frequency of the cantilever can be obtained by measuring the amplitude-frequency spectrum of vibrating cantilever using optical techniques. Figure 2(c) shows amplitude-frequency spectrum of a vibrating cantilever for three different conditions: before functionalizing with receptor molecules (blue circles), after functionalizing (red squares), and after capture of target molecules (black diamond) [19]. By definition, the peak in the amplitude-frequency spectrum corresponds to the resonance frequency $f_0$. As expected, resonance frequency decreases following the attachment of receptor molecules due to the added mass on the cantilever. It decreases *further* after capture of target molecules by receptor molecules due to further increase in the mass. Change in the resonance frequency due to the adsorption of the molecules can be obtained using Eq. (2), and is given by-

$$\frac{\Delta f}{f_0} \approx -\frac{\Delta m}{2m} + \frac{\Delta k}{2k}, \qquad (3)$$

where $\Delta m$ is the mass of added molecules and $\Delta k$ is the change in stiffness. Figure 2(d) shows $\Delta f$ as a function of $\Delta m$ for two different cantilevers [20] and linear dependence of $\Delta f$ on $\Delta m$ confirms Eq. 3 (assuming $\Delta k = 0$). Equation (3) suggests that the sensitivity $S \equiv \Delta f/f_0$ of nanocantilever biosensors can only vary linearly with $\Delta m$ (assuming $\Delta k = 0$) and therefore $S \propto N_s$ with $N_s$ being the areal density of captured bio-molecules on cantilever surface. We emphasize that these biosensors – with careful design and appropriate instrumentation – can be extraordinarily sensitive; indeed, zeptogram mass detection has been reported [21]. It is also important to realize that the linear sensitivity with $\Delta m$ is achieved only if the change in stiffness ($\Delta k$) due to capture of bio-molecules is negligible (Eq. (3)). In general, the capture of target molecules increases $k$ [19]. If $\Delta k$ compensates $\Delta m$, Eq. (3) suggests that there may be no change in resonance frequency at all (i.e., $\Delta f \sim 0$) and the sensitivity could be vanishingly small. Therefore, one must independently measure the change in $k$ to decouple the 'mass effect' from 'stiffness effect', so that the mass of the adsorbed molecule can be correctly estimated [22], [23].

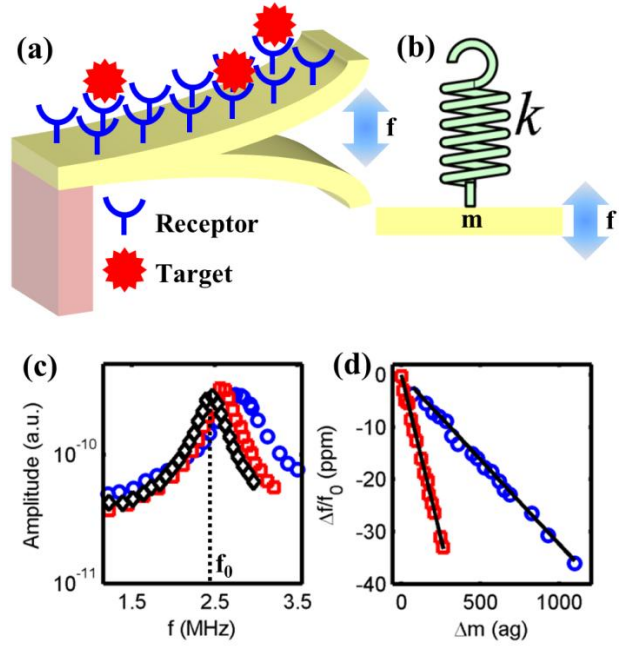

Fig. 2: Dynamic resonant mode sensing using nanocantilever. **(a)** Schematic of a vibrating cantilever whose resonance frequency changes due to capture of target bio-molecules. **(b)** Equivalent spring-mass model of the vibrating cantilever. **(c)** Amplitude vs. frequency spectrum of a free standing cantilever (○), after receptor molecules attachment (□), and after capture of target molecules (◊) [19]. **(d)** Change in resonance frequency as a function of added mass of the bio-molecules for two different cantilevers [20].

The sharpness of the peak (or the width of the amplitude-frequency spectrum) vibrating cantilever is characterized by its quality factor ($Q$) (Eq. (1)) and depends on the damping due to the surrounding medium. As $Q$ increases, resonance peak becomes sharper and width of the spectrum is reduced. Unfortunately, value of minimum detectable $\Delta f$ increases as $Q$ is reduced [24]. Therefore, measurements in vacuum or air can be more sensitive (capable of resolving small $\Delta f$ and therefore smaller $\Delta m$) as compared that in fluidic environment [24]. This dependence of $\Delta f$ on $Q$ has inspired design of suspended microchannel resonators [25–27] that do not suffer from $Q$ degradation due to the surrounding fluid. Such resonators are operated either in vacuum or in air and the fluid containing the target bio-molecules flows through the microchannel, embedded within the cantilever itself.

To summarize, the response of resonant mode nanocantilever based biosensors is linear with respect to the added mass of bio-molecules. Minimum detectable mass depends on the quality factor of the vibrating cantilever. And, suspended microchannel resonators can detect lower masses due to their high quality factors.

### 3.2 *Stress based static mode sensors*

Another class of nanocantilever sensor involves operation in the static mode, in which capture of target molecules introduces a surface stress [5]. Changes in the surface stress can be the result of an adsorption process or electrostatic interactions between charged molecules on the surface or conformational changes of the adsorbed molecules. This change in the surface stress bends the cantilever as shown in Fig. 3(a). The deflection of the tip of the cantilever $\Delta y$ is then measured as a signature of bio-molecules capture. Stoney's equation [28] relates the deflection $\Delta y$ with the change in the surface stress $\Delta \sigma$ as follows-

$$\Delta y = \frac{3L^2(1-\nu)}{EH^2} \Delta \sigma, \qquad (4)$$

where $L$ is the length, $\nu$ is the Poisson's ratio, $E$ is the Young's modulus, and $H$ is the thickness of the cantilever. Note that, Eq. (4) can be obtained from Eq. (1) (with time derivatives and $k' = 0$) with appropriately chosen $F_{ext} = -\frac{3\alpha W H}{L} \Delta \sigma$.

Deflection $\Delta y$ of the cantilever tip can be measured optically (e.g., using a laser and photodiode) or electrically (e.g. using an integrated piezo-resistor). Figure 3(b) shows deflection $\Delta y$ as a function of the target bio-molecules concentration in the solution for two different cantilevers having different geometrical dimensions [29]. The response $\Delta y$ is *sub-linear* with respect to the concentration and it depends on the geometrical dimensions of the cantilever. Using Eq. (4) and data shown in Fig. 3(b), $\Delta \sigma$ can be calculated and is shown in Fig. 3(c). Interestingly, $\Delta \sigma$ for two different cantilevers follow a single curve (Fig. 3(c)), suggesting that $\Delta \sigma$ only depends on the concentration of the bio-molecules and not on the cantilever properties.

Instead of optical measurement of $\Delta y$ or $\Delta \sigma$ through laser-photodiode system, one can measure $\Delta \sigma$ by measuring the change in the resistance of a piezoresistor attached to the cantilever [30]. For these piezoresistive based cantilever biosensors, the sensitivity is defined as the ratio of resistance after ($R$) and before ($R_0$) the capture of bio-molecules i.e., $S_{Piezo} \equiv R/R_0$. Figure 4(d) shows $S_{Piezo}$ for two different cantilevers, suggesting that response $S_{Piezo}$ only varies logarithmically with bio-molecules concentration. We, therefore, conclude that these static mode nanocantilever sensors respond only sub-linearly i.e., $S \sim \ln(N_s)$ (Figs. 3(b)-(d)) to target analyte concentration (Figs. 3(b)-(d)).

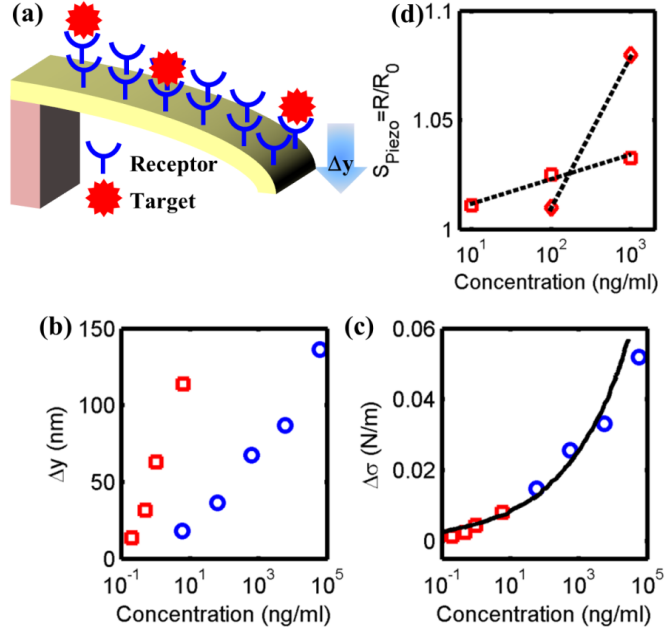

Fig. 3: Stress based sensing using the static response of a nanocantilever. **(a)** Schematic of a bent cantilever due to capture of bio-molecules on its surface. **(b)** Displacement of the tip of the cantilever as a function of bio-molecules concentration for two different cantilevers [29]. **(c)** Corresponding change in the surface stress of the cantilever. Symbols denote experimental data and solid line is just guide to the eye. **(d)** Ratio of the resistance of piezoresistive material attached to the cantilever after ($R$) and before ($R_0$) capture of bio-molecules [30]. Symbols denote experimental data and dotted line is just guide to the eye.

## 4. EMERGING NONLINEAR BIOSENSORS

In the previous section, we have discussed classical linear biosensors that can either be operated in static or dynamic mode. Now, we discuss a new class of emerging nonlinear biosensors that utilize inherent instability of nanocantilever static/dynamic response to achieve better sensitivity towards bio-molecules capture.

### 4.1 *Bifurcation based mass sensors*

As discussed above, classical resonant mode biosensors rely on the change in resonance frequency due to capture of bio-molecules. Note that, when operated in the linear regime (under small amplitude limit), the amplitude-frequency spectrum is symmetric and bell-shaped, as shown in Fig. 4(a). In this case, detection of bio-molecules is achieved by observing the shift in the peak (i.e., $\Delta f$, see Fig. 4(a)), as discussed in Sec 3.1.

In the large amplitude nonlinear response regime, however, higher order spring nonlinearities ($i.e., k' \neq 0$ in Eq. (1)) distorts the response, and amplitude-frequency spectrum is no longer symmetric [31]. Figure 4(b) shows one such amplitude-frequency spectrum with softening nonlinearity ($k' < 0$ and $F_{ext} = F_0 \sin(2\pi f_{ex} t)$ in Eq. (1) with $F_0$ being the excitation amplitude and $f_{ex}$ is the excitation frequency) for a Duffing like resonator. Interestingly, spectrum exhibits sudden jumps at points $P$ and $S$ representing saddle-node bifurcations. The hysteretic behavior shown in Fig. 4(b) is achieved, when $F_0 > F_c$ with $F_c$ being a critical threshold. Kumar et al., has proposed a bifurcation based mass sensor that utilizes these sudden jumps and rely on the shift in the amplitude and not on the shift in the frequency to signal bio-molecules capture [32]. In bifurcation based sensing, the resonator is operated near one of the critical point (say $P$). Capture of the bio-molecules reduces the fundamental frequency $f_0$ and increases $f_{ex}/f_0$ resulting in the sudden change in the amplitude of oscillation $\Delta A$, as shown in Figs. 4(b)-(c). Measurement of $\Delta A$ (using laser Doppler vibrometer (LDV)) is then used as the signature of capture of bio-molecules. It should be appreciated that this sensing scheme is very sensitive to small quantities of added molecules due to the amplification offered by inherent instability of mechanical system.

### 4.2 *Electromechanical coupling based Flexure-FET biosensors*

We have discussed both linear and nonlinear cantilever based biosensors that rely on optical readout of $y$ or $f_0$. Now, we discuss a new class of nonlinear biosensors called Flexure-FET [16] that utilize the electromechanical coupling between a suspended beam and a field effect transistor to achieve much higher sensitivity compared to traditional biosensors. Flexure-FET consists of a channel biased through a thin-film suspended gate (Fig. 5(a)). While the structure is similar to that of a suspended-gate FET [33], NEMFET [10] or resonant gate transistor [34], we call the sensor Flexure-FET to emphasize its distinctive nonlinear operation specifically optimized for ultrasensitive detection of bio-molecules. In a Flexure-FET, any change in the mechanical property of the suspended gate is directly reflected in the change of drain current of underneath field effect transistor and thereby enables electrical readout. As shown in Fig. 5(b), ultra high sensitivity arises from the coupling of two nonlinear responses, namely (i) spring-softening [35] in which stiffness decreases nonlinearly with the applied gate bias $V_G$ and vanishes at the pull-in point (for detailed discussions on pull-in instability, see Ref. [36], [37]), and (ii) sub-threshold electrical conduction [38] in which current depends exponentially on the surface potential. Such nonlinear electro-mechanical coupling enables exponentially high sensitivity for Flexure-FET sensors, which is fundamentally unachievable by exclusive use of existing nanoscale electronic or mechanical biosensors.

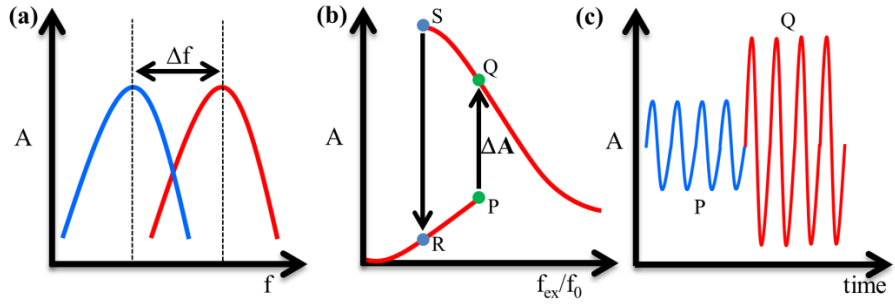

Fig. 4: Comparison of classical linear mass sensors with nonlinear bifurcation based mass sensors. Amplitude-frequency spectrum of **(a)** classical linear and **(b)** bifurcation based mass sensors proposed in ref. [32]. **(c)** Amplitude as a function of time in bifurcation based mass sensors. Classical sensors rely on the change in frequency $\Delta f$ whereas bifurcation based sensors rely on change in the amplitude $\Delta A$ due to capture of bio-molecules.

It should be noted that from a mechanical perspective, Flexure-FET operates close to pull-in instability, a critical point. Similar critical point sensing has also been reported for vapor sensors that operates close to bucking-instability [39], [40] and for mass sensor that operates close to saddle-node bifurcation [32] (discussed in Sec. 4.1) and their higher sensitivity have been confirmed experimentally. However, beyond the critical point sensing, the integrated transistor-action in the sub-threshold regime provides the Flexure-FET an *additional* exponential sensitivity (and simpler DC read-out) that could not be achieved by the classical nonlinear sensor schemes.

The operating principle of Flexure-FET can be understood based on a spring-mass system coupled to electrostatic actuation, see Fig. 6 [10], [34]. With the application of gate bias $V_G$, the gate moves downward towards the dielectric ($y$ $vs.$ $V_G$ curve in Fig. 5(b)) and the corresponding increase in gate capacitance is reflected in the increased drain current $I_{DS}$, as shown in Fig. 5(b). The static behavior of the device is dictated by the balance of spring and electrostatic forces (Eq. (1) with time derivatives zero and $k' = 0$), i.e.,

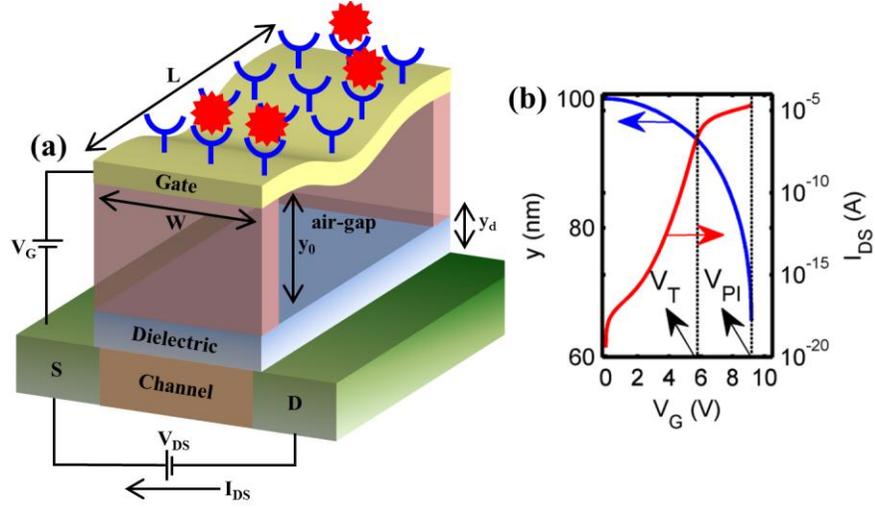

Fig. 5: Static stiffness based nonlinear electromechanical sensing with electrical readout. (a) Schematic of Flexure-FET bio-sensor and (b) $y$ and $I_{DS}$ vs. gate voltage characteristic of Flexure-FET. $V_T$ is the threshold voltage and $V_{PI}$ is the pull-in voltage of Flexure-FET.

$$k(y_0 - y) = F_e \equiv \frac{1}{2}\epsilon_0 E_{air}^2 A, \qquad (5)$$

where $E_{air}$ is the electric field in the air, and $A = WL$ is the area of the gate electrode. The electric field below the membrane $E_{air}$ is equal to $\epsilon_s E_s(\psi_s)$, where, $\epsilon_s$ is the dielectric constant of the substrate, and

$$E_s(\psi_s) = \sqrt{\frac{2qN_A}{\epsilon_0 \epsilon_s}} \left[ \psi_s + \left(e^{-\frac{q\psi_s}{k_B T}} - 1\right)\frac{k_B T}{q} \right.$$
$$\left. - \left(\frac{n_i}{N_A}\right)^2 \left(\psi_s - \left(e^{\frac{q\psi_s}{k_B T}} - 1\right)\frac{k_B T}{q}\right)\right]^{\frac{1}{2}}, \qquad (6a)$$

where, $E_s(\psi_s)$ is the electric field at the substrate-dielectric interface (see Ref. [38] page 64 for a detailed derivation of Eq. 6(a)), $\psi_s$ is the surface potential, $q$ is the charge of an electron, $N_A$ is the substrate doping, $k_B$ is the Boltzmann constant, $T$ is the absolute temperature, and $n_i$ is the intrinsic carrier concentration in the substrate. The voltage drop in air ($y\epsilon_s E_s(\psi_s)$), dielectric $\left(\frac{y_d}{\epsilon_d}\epsilon_s E_s(\psi_s)\right)$, and substrate ($\psi_s$) can be related to the applied gate bias $V_G$ as follows-

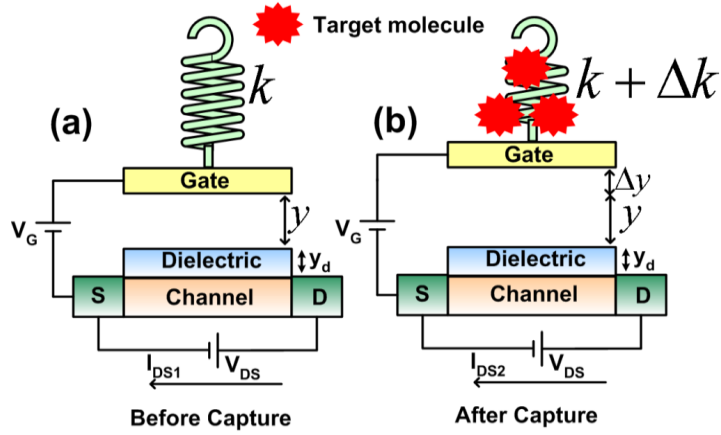

Fig. 6 (a)-(b) Equivalent spring-mass model of Flexure-FET. Stiffness changes from $k$ to $k + \Delta k$ after the capture of biomolecules, and therefore, position of gate changes from $y$ to $y + \Delta y$ which results in the modulation of drain current from $I_{DS1}$ to $I_{DS2}$.

$$V_G = \left(y + \frac{y_d}{\epsilon_d}\right)\epsilon_s E_s(\psi_s) + \psi_s, \tag{6b}$$

where, $y_d$ is the dielectric thickness. Equations (5)-(6) are solved self-consistently for $y$ and $\psi_s$ at each $V_G$. The corresponding inversion charge density ($Q_i$) in the channel and drain current ($I_{DS}$) are given by,

$$Q_i = \frac{qn_i^2}{N_A}\int_0^{\psi_s} \frac{e^{\frac{q\psi}{k_B T}} - 1}{E_s(\psi)} d\psi, \tag{7}$$

$$I_{DS} = \mu_n L Q_i \frac{V_{DS}}{W}, \tag{8}$$

where, $\mu_n$ is the channel mobility for electrons, $V_{DS}$ is the applied drain to source voltage. Figure 5(b) shows the steady-state response of Flexure-FET as a function of biasing voltage $V_G$, obtained from the numerical simulations of Eqs. (5)-(8).

For transduction, Flexure-FET biosensor utilizes the change in suspended-gate stiffness from $k$ to $k + \Delta k$, due to the capture of bio-molecules. The change in stiffness due to the capture of bio-molecules has been demonstrated by several recent experiments of mass sensing using nanocantilever based resonators [19], [41–43]. This well-known observation of stiffness change has been attributed to the change in the membrane thickness, Young's modulus, and/or surface stress of the beam [19], [41–43]. Indeed, Craighead in Ref. [44] suggests its use as a basis of a new class of mechanical biosensor.

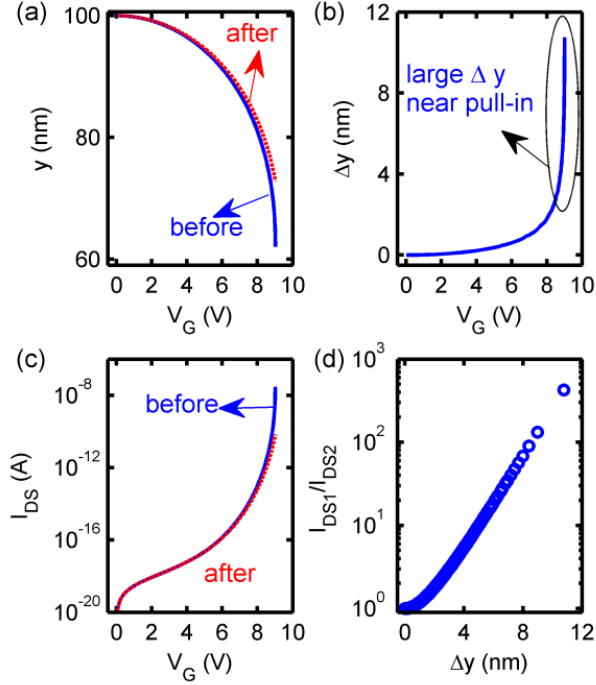

Fig. 7: Change in the sensor characteristics due to capture of target molecules on the surface of the gate, (a) $y$ vs. $V_G$ before and after capture, and (b) corresponding change in the position of gate electrode $\Delta y$ vs. $V_G$. $\Delta y$ increases rapidly near pull-in due to spring-softening effect. The capture of target molecules is directly mirrored in the change in $I_{DS}$. (c) $I_{DS}$ vs. $V_G$ for before and after capture, and (d) corresponding ratio of the two currents $I_{DS1}$ (before capture) and $I_{DS2}$ (after capture) as a function of $\Delta y$. Orders of magnitude change in $I_{DS}$ can be easily achieved for typical surface density of $N_s = 5*10^{12} cm^{-2}$, projected area of the bio-molecule, $A_t = \pi R_t^2$ with $R_t = 1nm$, and $H_t = 5.1nm$. These parameters translate to just an equivalent $\Delta k \sim 6\%$. The device considered has the following typical parameters: $L = 4\mu m, W = 1\mu m, H = 40nm, E = 200GPa, y_0 = 100nm, y_d = 5nm, \epsilon_s = 11.7, \epsilon_d = 3.9, N_A = 6*10^{16} cm^{-3}$.

In the following analysis, we model change in $k$ by the change in the effective thickness $H$ of the gate ($\Delta H$), *although it should be stressed that the conclusions do not depend on the particular hypothesis regarding $\Delta k$*. For now, we ignore the details of the spatial distribution of molecules associated with random sequential adsorption [15], and assume a uniform distribution of adsorbed molecules on the sensor surface. Therefore, the conservation of volume suggests $\Delta H = N_s A_t H_t$, where $N_s$ is the areal density, $A_t$ is the effective cross-sectional area, and $H_t$ is the effective thickness of the target molecule. Using the fact that $k = \frac{\alpha E W H^3}{(1-\nu)L^3}$, change in stiffness $\Delta k$ due to $\Delta H$ ($\ll H$) can be related to adsorbed molecule density $N_s$ as follows,

$$\frac{\Delta k}{k} \approx \frac{3 N_s A_t H_t}{H}. \qquad (9)$$

It can be shown that if Flexure-FET is operated close to pull-in and in sub-threshold regime, sensitivity $S$ (using Eqs. (5)-(9), see Ref. [16] for details of derivation) is given by-

$$S_{Flexure} \equiv \frac{I_{DS1}}{I_{DS2}} \approx \exp(\gamma_1 \sqrt{N_s} - \gamma_2 N_s), \qquad (10)$$

where $\gamma_1/\gamma_2$ are two sensor geometry dependent constant. Equation 10 confirms the exponential sensitivity of Flexure-FET towards bio-molecules capture.

The results for the change in sensor characteristics due to the capture of bio-molecules are summarized in Fig. 7. For example, Fig. 7(a) shows $y$ vs. $V_G$ before and after capture of target molecules. After the capture, gate moves up (for a fixed $V_G$) due to increased restoring spring force (because of increase in $k$, see Fig. 7(a)). Interestingly, change in gate position $\Delta y$ is maximum close to pull-in due to spring-softening effect, as shown in Fig. 7(b). The change in gate position $\Delta y$ is directly reflected in change in $I_{DS}$. Figure 7(c) shows $I_{DS}$ vs. $V_G$ before and after capture of bio-molecules. Interestingly, $I_{DS}$ *decreases* after capture due to increased separation between the gate and the dielectric (hence decreased capacitance). The corresponding ratio of the currents $I_{DS1}$ (before capture) and $I_{DS2}$ (after capture) increases exponentially with $\Delta y$ (Fig. 7(d)), and becomes maximum near pull-in.

## 5. CONCLUSIONS

In this review article, we have discussed various ways of detecting bio-molecules using nanocantilevers. Classical resonant mode biosensors detect the change in the resonance frequency of vibrating cantilever and require complex optical instrumentation for detection, especially when very high sensitivity is desired. Stress based static mode sensors detect the deflection of the tip of the cantilever and responds linearly to change in the stress. To achieve better sensitivity than achieved by classical linear biosensors, critical-point nonlinear bio-sensors have started to appear in the literature. For example, we discussed bifurcation based mass sensors that operate close to a saddle-node bifurcation. Finally, we have discussed Flexure-FET biosensor that integrates a transistor for direct electrical readout and utilizes nonlinear electromechanical coupling for its exponential sensitivity. We believe that these critical point nonlinear biosensors with electrical readout will offer opportunity to integrate highly sensitive sensors in low cost point-of-care applications.